\documentclass[aps,prb,twocolumn,groupedaddress,amsmath,amssymb,amsfonts,graphics]{revtex4}

\newcommand{\bracket}[1]{\langle #1 \rangle}
\newcommand{\ket}[1]{\rvert #1 \rangle}
\newcommand{\bra}[1]{\langle #1 \lvert}
\usepackage{graphicx}
\usepackage{bm}
\usepackage{bbold}
\newcommand{\svec}[1]{\bm{#1}}
\renewcommand{\vec}[1]{\mathbf{#1}}

\begin{document}

\title{Master equation approach to computing RVB bond amplitudes}

\author{K.\ S.\ D.\ Beach}
\affiliation{Institut f\"{u}r Theoretische Physik,
Universit\"{a}t W\"{u}rzburg, Am Hubland, 97074
W\"{u}rzburg, Germany}

\date{September 20, 2007}

\begin{abstract}
We describe a ``master equation'' analysis for the bond amplitudes $h(\vec{r})$ of an RVB 
wavefunction. Starting from any initial guess, $h(\vec{r})$ evolves---in a manner dictated
by the spin hamiltonian under consideration---toward a steady-state 
distribution representing an approximation to the true ground state. 
Unknown transition coefficients in the master equation are treated as variational parameters.
We illustrate the method by applying it to the $J_1$--$J_2$ antiferromagnetic Heisenberg
model. Without frustration ($J_2=0$), the amplitudes are radially symmetric and fall off as $1/r^3$ 
in the bond length. As the frustration increases, there are precursor signs of 
columnar or plaquette VBS order: the bonds preferentially align along the axes of
the square lattice and weight accrues in the nearest-neighbour bond amplitudes.
The Marshall sign rule holds over a large range of couplings, $J_2/J_1 \lesssim 0.418$.
It fails when the $\vec{r}=(2,1)$ bond amplitude first goes negative, a point also marked
by a cusp in the ground state energy. A nonrigourous extrapolation of the
staggered magnetic moment (through this point of nonanalyticity) shows it
vanishing continuously at a critical value $J_2/J_1 \approx 0.447$. This
may be preempted by a first-order transition to a state of broken translational symmetry.
\end{abstract}

\maketitle

\section{Introduction}

In the early 1970s, a resonating-valence-bond (RVB) wavefunction~\cite{Pauling49} 
with nearest-neighbour (NN) bonds only was proposed as a possible ground state for the 
quantum Heisenberg model on the triangular lattice.~\cite{Anderson73,Fazekas74} 
This short-ranged,  quantum-disordered~\cite{Rokhsar88,Moessner01} RVB state
was conceived in analogy with the spin liquid state found in one dimension.~\cite{Bethe31,Hulthen38}
The belief was that classical 120$^\circ$ N\'{e}el order was unlikely to survive 
in the presence of strong quantum fluctuations.

This conjecture ultimately proved incorrect. Like other low-coordination-number 
antiferromagnets,~\cite{Casto06, Sandvik97} the triangular system
is ordered at zero temperature.~\cite{Huse88,Singh92,Capriotti99}
Consequently, its ground state cannot be described in a basis of short bonds.
One can show, in fact, that a correct description must involve valence bonds on all length 
scales.~\cite{NoteA}

A generalization of the RVB state that includes long bonds was later
proposed by Liang, Doucot, and Anderson for use as a variational wavefunction
in the square-lattice Heisenberg model.~\cite{Liang88} Their idea was 
to factorize the weight associated with each valence bond configuration 
into a product of individual bond amplitudes that depend only on the
vector $\vec{r}$ connecting bond endpoints.
Unlike the NN-bond RVB, which is unique, the long-range version is a family of states 
parameterized by the bond distribution function, $h(\vec{r})$. 
In $d=2$, the RVB wavefunction has expressive power to
describe both an antiferromagnetically ordered phase 
and a featureless quantum disordered phase.~\cite{Liang88,Beach07b} 
It may be a good variational wavefunction for systems in which antiferromagnetism 
is killed by the addition of frustrating interactions.

As a practical matter, optimizing the bond amplitudes numerically is
not straightforward. The number of independent parameters is of the 
order of the system size, and the energy depends only very weakly 
on the amplitudes of the longest bonds. Thus, obtaining well-converged
results becomes increasingly difficult for large lattices, and scaling to the 
thermodynamic limit is unreliable. Lou and Sandvik~\cite{Lou06} have made
some progress by experimenting with different optimization schemes.
They recently carried out an unbiased variational determination of 
$h(\vec{r})$ for the square-lattice Heisenberg model and were able to
achieve lattice sizes up to $32 \times 32$.

Liang, Doucot, and Anderson circumvent the problems associated with
a macroscopic number of degrees of freedom by assuming a functional 
form for $h(\vec{r})$. They vary the amplitudes of only a few short bonds 
and fix the remainder under the assumption of a radially symmetric 
bond-length distribution and algebraic decay at long distances.~\cite{Liang88}
For local, nonfrustrating interactions, this assumption turns out to
be essentially correct.~\cite{NoteB} Nonetheless, their choice of functional 
form is \emph{ad hoc}, and there is nothing in their approach that provides
insight into how the amplitudes might change when competing interactions are introduced.

In this paper, we describe an alternative method for calculating 
the bond amplitudes that requires at most a few variational parameters.
The utility of the method is tested by applying it to the $J_1$--$J_2$ model.
As in Ref.~\onlinecite{Liang88}, we make strong assumptions about 
the form of the bond distribution. In our case, however, the choice of
functional form for $h(\vec{r})$ is guided by a master equation that 
mimics the reconfiguration of bond amplitudes induced by the evolution
operator.

The $J_1$--$J_2$ model describes a system of spin-half moments arranged
on a square lattice in which Heisenberg interactions of strength $J_1$, acting along 
the plaquette edges, compete with frustrating interactions of strengh 
$J_2$, acting across the plaquette diagonals. At $J_2/J_1=0$ and 
$J_2/J_1=\infty$, the model has two- and four-sublattice N\'{e}el order, 
respectively. There is a gapped intermediate phase in the vicinity of 
$J_2/J_1 \approx 0.5$, whose exact nature remains controversial.
There has been speculation about a possible spin liquid
state,~\cite{Chandra88,Figueiriedo89,Oguchi90,Schulz92,Zhang03}
but a state with broken translational symmetry  now seems more likely.
The leading candidate is a valence bond solid (VBS) with either 
columnar~\cite{Dagotto89,Gelfand89,Gelfand90,Singh90}
or plaquette~\cite{Zhitomirsky96,Capriotti00,Mambrini06} order.

The extent of the intermediate phase has been determined to
about one digit of precision. Exact diagonalization on small 
clusters~\cite{Schulz96} puts the lower critical point at $J_2/J_1 = 0.34(4)$, 
but this appears to be an underestimate. Bond operator calculations~\cite{Kotov99,Kotov00a} 
based on the columnar VBS predict $0.38 \lesssim J_2/J_1\lesssim 0.62$ for the region
of stability, and series expansions~\cite{Oitmaa96,Singh99} from the magnetic side 
give $0.4 \lesssim J_2/J_1 \lesssim 0.6$.
A quantum Monte Carlo study,~\cite{{Sorella98}} in which stochastic reconfiguration
is used to partially alleviate the sign problem, reports a transition to a gapped state
at $J_2/J_1 \approx 0.4$.

It has been established from energy level crossings in series expansion
that the transition at the upper critical point is first order.~\cite{Kotov00b}
No such crossings have been detected at the lower critical point, 
at least within the numerical accuracy that can be achieved.
In most of the studies cited above, it is implicitly assumed that the
transition at the lower critical point is second order. If that is true---and 
if the intermediate phase is indeed bond ordered---then the lower critical
point may constitute a deconfined quantum critical point, as envisioned 
by Senthil \emph{et al}.~\cite{Senthil04}

The fact that bond operator methods indicate a high density of triplet modes 
near a deconfined quantum critical point,~\cite{Kotov07} but only a low density
near the critical point of the $J_1$--$J_2$ model,~\cite{Kotov99}
leaves room for doubt. Indeed, a recent series expansion 
study points to a \emph{first order} transition at $J_2/J_1 \approx 0.43$
on the basis of an energy functional computed for a 
fictitious translational-symmetry-breaking field.~\cite{Sirker06} 
This is supported by the argument due to Chubukov~\cite{Chubukov91} 
that a continuous transition is only possible when a third-nearest-neighbour 
interaction $J_3 > 0$ is present.

The results reported here cannot settle this question
with any certainty, but they do appear to be more
consistent with a first order N\'{e}el--VBS transition.

\section{Master equation for factorizable RVB bond amplitudes}

\begin{figure}
\includegraphics{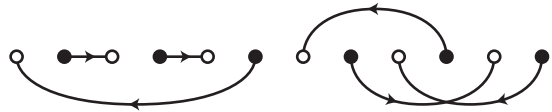}
\caption{\label{FIG:vb}Spins on the A sublattice (solid circles) and B sublattice (open circles) are grouped
into pairs forming singlets. Each pairing configuration is characterized by a set of directed bonds connecting A sites to B sites.}
\end{figure}

The spin-rotation-invariant (total spin $S=0$) ground state of a system of
$2N$ spin-$\tfrac{1}{2}$ moments can
be written as a superposition of valence bond states.~\cite{Rumer32,Pauling33}
The simplest RVB ansatz is to assume that the weight given to each bond configuration
is a product of individual bond amplitudes:
\begin{equation} \label{EQ:RVBwf}
\lvert h \rangle = \sum_{v} \biggl[ \prod h(\vec{r}) \biggr] \lvert v \rangle.
\end{equation}
Here the sum is over all partitions of the lattice into $N$ singlet pairs,
and the product is over all vectors $\vec{r}$ drawn between valence bond endpoints.
[Anderson's NN-bond RVB corresponds to $h(\vec{r})
= \delta(\lvert\vec{r}\rvert-1)$.]
In the special case of a nonfrustrated model on a bipartite lattice,
the amplitudes $h(\vec{r})$ are real and nonnegative and 
strictly zero whenever $\vec{r}$ connects valence bonds in the same sublattice.
See Fig.~\ref{FIG:vb}. This is just a restatement of the Marshall sign theorem.~\cite{Marshall55}

One way to compute the $h(\vec{r})$ values appropriate for a given model is 
to consider the $\tau$-dependent family of states
\begin{equation} \label{EQ:htau}
\lvert h(\tau) \rangle = e^{-\tau \hat{\mathcal{F}}\hat{H}\hat{\mathcal{F}} } \lvert h(0) \rangle,
\end{equation}
where $\hat{H}$ is the hamiltonian of interest and $\hat{\mathcal{F}}$ is an operator that 
projects onto the space of factorizable RVB wavefunctions. 
In each time step $d\tau$, some fraction of the bond amplitude is reapportioned as 
bonds are created and destroyed.
Correlations between bonds that go beyond the RVB framework are prevented from accumulating.
This process is governed by a master equation that 
describes how the distribution $h(\vec{r})$ 
evolves towards its steady-state solution.
Note that the wavefunction that emerges in the $\tau \to \infty$ limit
is not strictly equal to the projection $\hat{\mathcal{F}} \ket{\psi}$ of the true ground
state $\ket{\psi}$; nor is it equal to the variationally determined state 
$\ket{h}$ that minimizes $E = \langle h \lvert \hat{H} \rvert h \rangle/\langle h | h \rangle$.
Nonetheless, all three are very similar to one another.~\cite{Sandvik07b}

\begin{figure}
\includegraphics{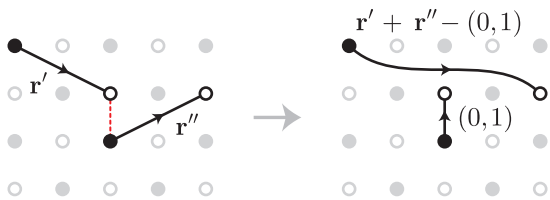}
\caption{\label{FIG:updatesAB}
A nonfrustrating Heisenberg interaction, indicated by the dotted (red) line,
is applied between sites in opposite sublattices. The resulting
reconfiguration creates one valence bond where the interaction
was applied and another between the two remaining endpoints.
}
\end{figure}

The key observation is that the valence bond basis is closed under operation by the Heisenberg interaction.
Operating on an existing bond simply leaves the bond as is [and the distribution $h(\vec{r})$
unchanged] whereas operating between two bonds maps them to their complementary tiling:
\begin{align} \label{EQ:updaterule1}
\Bigl(\frac{1}{4}-\mathbf{S}_i\cdot\mathbf{S}_j\Bigr)[i,j] &= [i,j],\\ \label{EQ:updaterule2}
\Bigl(\frac{1}{4}-\mathbf{S}_i\cdot\mathbf{S}_j\Bigr)[i,l][k,j] &= \frac{1}{2}[i,j][k,l].
\end{align}
Here $[i,j] = \frac{1}{\sqrt{2}}( \lvert \uparrow_i \downarrow_j\rangle
-\lvert \downarrow_i \uparrow_j\rangle$ denotes a singlet formed from
the spins at sites $i$ and $j$.
The effect of Eq.~\eqref{EQ:updaterule2} is depicted in Fig.~\ref{FIG:updatesAB}. 

For the NN Heisenberg model on a $d$-dimensional (hyper-)cubic lattice,
the master equation is
\begin{equation} \label{EQ:Master}
\dot{h}(\vec{r}) = \sum_{\vec{a}} \bigl[\delta_{\vec{r},\vec{a}}  + \sum_{\vec{r}',\vec{r}''}
\delta_{\vec{r'}+\vec{r}'' - \vec{a},\vec{r}}h(\vec{r'}) h(\vec{r}'')\bigr] - 2zh(\vec{r}),
\end{equation}
where $\dot{h} = \partial h/\partial\tau$, $z = 2d$ is the coordination number, and $\vec{a}$ ranges over all 
NN vectors. This is correct only insofar as $h(\vec{r})$
accurately measures how often a bond of type $\vec{r}$ appears 
in the superposition of valence bond configurations making up the RVB state. 
Geometrical tiling constraints, which are increasingly important at low
coordination number, have been ignored. Nonetheless, this level of
approximation allows us to proceed analytically.

Equation~\eqref{EQ:Master} conserves the
unit normalization of the total weight:
\begin{equation}
\sum_{\vec{r}} \dot{h}(\vec{r}) = z + z\biggl[\sum_{\vec{r}} h(\vec{r}) \biggr]^2 
- 2z \sum_{\vec{r}} h(\vec{r}) = 0.
\end{equation}
The AB character of the bonds is also a constant of the motion.
If we start with a distribution $h(\vec{r})$ that is nonzero only when $\vec{r}$ 
connects sites in opposite sublattices, then $h(\vec{r})$
will also have this property at all subsequent $\tau$. 

A somewhat stronger property of the flow is that all weights associated with 
bonds of even Manhattan length $\lVert \vec{r} \rVert = |r_1| + |r_2| + \cdots |r_d|$,
namely the AA or BB bonds, are driven to zero. This is a straightforward 
consequence of an asymmetry in the reconfiguration rules:
(even,odd)$\,\to\,$(even,odd), (odd,odd)$\,\to\,$(odd,odd), and
(even,even)$\,\to\,$(odd,odd).
This is yet another manifestation of the Marshall sign rule.

\begin{figure}
\includegraphics{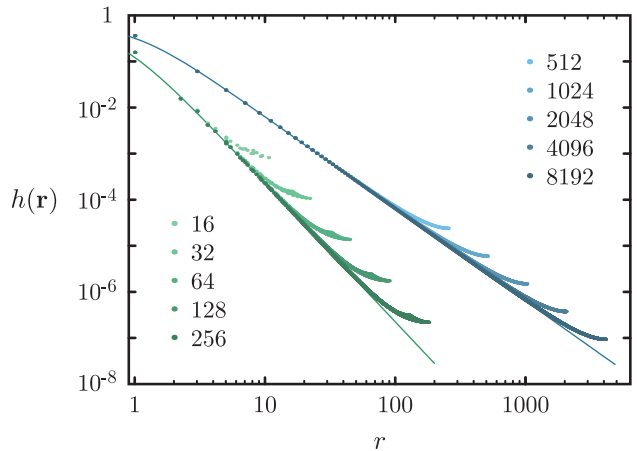}
\caption{\label{FIG:h1and2d} The bond amplitude functions predicted for the
linear-chain (upper, blue curves) and square-lattice (lower, green curves)
Heisenberg models are plotted for various system sizes. 
Larger values of $L$ are indicated by darker data points, following the legend. 
The solid lines reflect the analytical results given in Eqs.~\eqref{EQ:h1d}
and \eqref{EQ:h2d}.
}
\end{figure}

Accordingly, for $\tau \to \infty$ there are no bonds connecting sites in the 
same sublattice and all bonds have odd Manhattan length. We are thus free to
impose the convention that the vector character of all bonds is directed from A to B
(as anticipated in Fig.~\ref{FIG:vb}).
This means that the bond amplitude function has a Fourier expansion
\begin{equation}
h(\vec{r}) =  \frac{1}{N}\sum_{\vec{q}} e^{i \vec{q}\cdot\vec{r}} h_{\vec{q}},
\end{equation}
where the wavevector sum ranges over a reduced ``magnetic'' Brillouin zone,
equal to the standard Wigner-Seitz cell modulo $\vec{Q} = (\pi, \ldots \pi)$.
One finds that the Fourier transform of Eq.~\eqref{EQ:Master} is a simple polynomial in $h_\vec{q}$,
\begin{equation}
\frac{1}{z}\dot{h}_{\vec{q}} = \gamma_{\vec{q}} + \gamma_{\vec{q}} h_{\vec{q}}^2 - 2h_{\vec{q}},
\end{equation}
whose stationary distribution is
\begin{equation} \label{EQ:hq}
h_{\vec{q}} = \frac{1-(1-\gamma_{\vec{q}}^2)^{1/2}}{\gamma_{\vec{q}}}.
\end{equation}
$\gamma_{\vec{q}} = (1/d)(\cos q_1 + \cdots \cos q_d)$ is the Fourier
transform of the  NN matrix.
In real space, the long distance behaviour is given by
\begin{alignat}{2} \label{EQ:h1d}
h(r) &= \frac{2}{\pi(1 + r^2)} & (d&=1) \\ \label{EQ:h2d}
h(\vec{r}) &= \frac{\sqrt{2}}{2\pi(\frac{1}{2} + r^2)^{3/2}} \qquad & (d&=2)
\end{alignat}
as shown in Fig.~\ref{FIG:h1and2d}. Note that in two dimensions, the bond amplitude
is almost prefectly radially symmetric beyond a few lattice spacings.
The general behaviour for  higher dimensions is $h(\vec{r}) 
\sim (1/d + r^2)^{-(d+1)/2} \sim r^{-(d+1)}$.

\section{\label{SECT:frustint}Frustrating interactions}

As we emphasized in the previous section, any model on a bipartite lattice
whose interactions are nonfrustrating with respect to two-sublattice N\'{e}el order
can be described in a basis consisting only of AB valence bonds.~\cite{Beach06,Alet07,Mambrini07}
Two special features of the AB basis are that (1) the overlap between any two 
states is strictly positive,~\cite{Sutherland88} and (2) there is an exact 
correspondence between the Marshall sign rule and the
positivity of all the RVB bond amplitudes.

\begin{figure}
\includegraphics{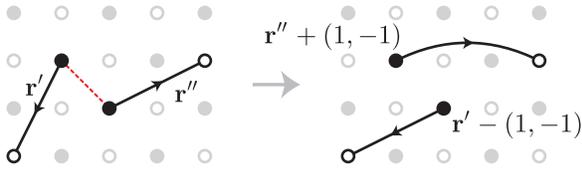}
\caption{\label{FIG:updatesAA}
A frustrating Heisenberg interaction applied between two A sublattice sites,
indicated by the dotted (red) line, has the effect of exchanging the two
valence bond endpoints.
}
\end{figure}

We now argue that, even with the addition of frustrating interactions,
one can still chose to work exclusively in the AB basis. 
According to Eq.~\eqref{EQ:updaterule1}, a frustrating interaction applied between sites 
in the same sublattice transforms two AB bonds into one AA and one BB bond. But since
valence bonds are nonorthogonal, we can take advantage of the overcompleteness relation
\begin{equation}
[i,k][j,l] = [i,j][k,l] - [i,l][k,j]
\end{equation}
to eliminate each of the unwanted bonds, yielding a new update rule
\begin{equation} \label{EQ:AAupdaterule}
\Bigl(\frac{1}{4}+\mathbf{S}_i\cdot\mathbf{S}_k\Bigr)[i,l][k,j] = \frac{1}{2}[i,j][k,l],
\end{equation}
where $i,k \in A$ and $j,l \in B$. See Fig.~\ref{FIG:updatesAA}.
There is no diagonal operation analogous to Eq.~\eqref{EQ:updaterule1}.

Following Eq.~\eqref{EQ:AAupdaterule}, a model with NN
Heisenberg interactions of strengh $J_1$ and next-nearest-neighbour 
(NNN) interactions of strength $J_2$
has the bond amplitude master equation
\begin{widetext}
\begin{equation} \label{EQ:J1J2mastereq}
\dot{h}(\vec{r}) = \sum_{\vec{a}} \bigl[\delta_{\vec{r},\vec{a}}  + \sum_{\vec{r}',\vec{r}''}
\delta_{\vec{r'}+\vec{r}'' - \vec{a},\vec{r}}h(\vec{r'}) h(\vec{r}'') \bigr] - 2zh(\vec{r})\\
+ \frac{J_2}{J_1} \biggl( \sum_{\tilde{\vec{a}}}\sum_{\vec{r}',\vec{r}''}\bigl[
\delta_{\vec{r}'-\tilde{\vec{a}},\vec{r}}
+ \delta_{\vec{r}''+\tilde{\vec{a}},\vec{r}}\bigr] h(\vec{r}')h(\vec{r}'')\\ - 2\tilde{z}h(\vec{r}) \biggr),
\end{equation}
\end{widetext}
where $\tilde{\vec{a}}$ ranges over all NNN vectors.
(We use a tilde to distinguish NNN quantities from NN ones.)
This differs from Eq.~\eqref{EQ:Master} by a term proportional to $J_2/J_1$.

Fourier transformation of Eq.~\eqref{EQ:J1J2mastereq} leads to
\begin{equation}
\frac{1}{z}\dot{h}_{\vec{q}} = \gamma_{\vec{q}} + \gamma_{\vec{q}} h_{\vec{q}}^2
-2\Bigl[ 1 + g\bigl( 1 - \tilde{\gamma}_\vec{q}\bigr)\Bigr] h_{\vec{q}}.
\end{equation}
This has a steady state solution
\begin{equation} \label{EQ:hqnnn}
h_{\vec{q}} = \frac{\Lambda_{\vec{q}} - (\Lambda_{\vec{q}}^2 - \gamma_{\vec{q}}^2)^{1/2}}{\gamma_{\vec{q}}},
\end{equation}
where
\begin{equation}
\Lambda_{\vec{q}} = 1+g(1-\tilde{\gamma}_{\vec{q}}) \ \ \text{and} \ \ g = (\tilde{z}/z)(J_2/J_1).
\end{equation}
In dimension $d>1$, the coordination numbers are
$z = 2d$, $\tilde{z} = 2^d$
and the connection matrices have Fourier transforms
$\gamma_{\vec{q}} = \tfrac{1}{d}\sum_{k=1}^d\cos q_k$
and $\tilde{\gamma}_{\vec{q}} = \prod_{k=1}^d\cos q_k$.

\begin{figure*}
\includegraphics{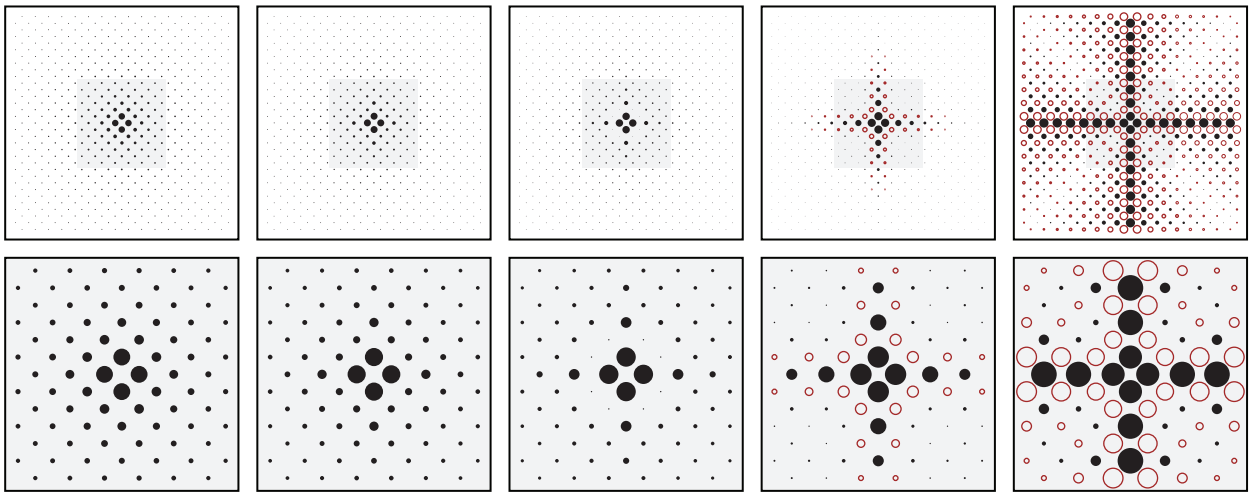}
\caption{\label{FIG:hr}The bond amplitudes $h(\vec{r})$ are depicted from left to right
for the values $g=0, 0.2, 0.32, 0.45, 0.49999$. 
The top row shows the subset of bonds up to length (16,15)
for an $L=256$ square lattice. The bottom row is a magnified view emphasizing 
the short range bonds up to (6,5).
The position of each circle marks the endpoint of a bond whose
other endpoint is at the origin. The area of each circle is proportional 
to $\lvert h(\vec{r}) \rvert r^{3/2}$.
Filled (black) circles denote a positive value and open (red) circles
a negative one.
For $g\le 0$, the bond amplitudes are radially symmetric
and positive definite and fall off as $r^{-3}$. As $g$ increases, the distribution 
becomes increasingly asymmetric. 
The (2,1) bond steadily decreases in magnitude and vanishes at $g_M = 0.323158$.
In the limit $g \to 0.5$, the bonds along the x and y axes become extremely
long ranged: $h(r,0) = h(0,r) \sim r^{-3/2}$.
For $g > 0.5$, the bond amplitudes are complex.
}
\end{figure*}

Figure~\ref{FIG:hr} illustrates the real-space distribution corresponding to 
Eq.~\eqref{EQ:hqnnn} for several values of $g$ in two dimensions. 
When $g\le 0$, the long-range behaviour
is $h(\vec{r}) \sim r^{-3}$, as in Eq.~\eqref{EQ:h2d}. When $g > 0$, the radial symmetry 
is reduced to the C${}_4$ symmetry of the square lattice,
and the amplitudes begin to accumulate along the principle axes,
especially in the $\vec{r} = (1,0)$ bond. Contrary to our expectations, 
the distribution does not become uniformly more short-ranged as $g$ increases.
Along the principle axes, it actually becomes longer-ranged: the exponent
of the algebraic decay steadily decreases from 3 (at $g=0$) to 1.5 (at $g=0.5$).
This looks nothing like the $h(\vec{r}) \sim r^{-p}$ spin liquid found at $p\gtrsim 3.3$.~\cite{Liang88,Beach07b}

The Marshall sign rule is obeyed in the $J_1$--$J_2$ model up to relatively 
large values of the frustration parameter.~\cite{Richter94,Ivanov94,Voigt97}
At the level of approximation employed here, the bond amplitudes are all 
positive up to $g_M =  0.323158$, the coupling
at which the amplitude of the $\vec{r}=(2,1)$ bond passes through zero.
A sign change in $h(2,1)$ at large frustration has also been observed 
by Lou and Sandvik in their unbiased calculation.~\cite{Lou06}

Since there is already some ambiguity in the master equation because
of the neglect of geometric constraints, we will treat $g$ as a variational
parameter. In other words, we will allow the relative weighting between the frustrating and nonfrustrating 
channels to deviate from $g=(\tilde{z}/z)(J_2/J_1)$, as the energy dictates.

In principle, the variational approach can be expanded to include
farther-neighbour moments, defined by
\begin{equation}
\gamma_{\vec{q}}^{\svec{\mu}} = \frac{1}{d!}\sum_{\sigma \in \mathcal{S}_d}
\prod_{n=1}^d \cos(\mu_{\sigma(n)}k_n).
\end{equation}
Here, the index $\svec{\mu}$ is an ordered $d$-tuple of natural numbers,
and $\mathcal{S}_d$ is the set of permutations on $d$ elements.
There will be a variational parameter $g^{\svec{\mu}}$ for each included moment,
in terms of which the amplitude distribution is 
\begin{equation}
h_{\vec{q}} = \frac{\lambda-\eta_{\vec{q}} - \bigl[(\lambda-\eta_{\vec{q}})^2 - \Delta_{\vec{q}}^2
\bigr]^{1/2}}{\Delta_{\vec{q}}},
\end{equation}
where
\begin{equation} \label{EQ:etadelta}
\eta_{\vec{q}} = \sum_{\text{even}} g^{\svec{\mu}}\gamma_{\vec{q}}^{\svec{\mu}}, \ \ \text{and} \ \ 
\Delta_{\vec{q}} = \sum_{\text{odd}} g^{\svec{\mu}}\gamma_{\vec{q}}^{\svec{\mu}}.
\end{equation}
$\lambda$ is fixed by the requirement that $h_{\vec{q}=0}=1$.
The summations in Eq.~\eqref{EQ:etadelta} are over all $\svec{\mu}$ vectors having even and odd Manhattan length
up to some cutoff, $\lVert \svec{\mu} \rVert < \mu_0$.
For our numerical work on the $J_1$--$J_2$ model, only the
$\svec{\mu} = (1,0)$ and $\svec{\mu} = (1,1)$ components are kept.

\section{Results for the $J_1$--$J_2$ model}

We work with an RVB trial wavefunction whose weights
are factorized as in Eq.~\eqref{EQ:RVBwf}. The bond amplitudes
are taken from the Fourier transform of Eq.~\eqref{EQ:hqnnn}.
These depend only on the size of the lattice
and on a single variational parameter, $g$, 
which is fixed by minimizing  
$E(g) = \bracket{\hat{H}}$. Expectation
values of an operator $\hat{O}$ in the trial state, written
\begin{equation}
\bracket{\hat{O}}\equiv
\frac{\bra{h}\hat{O}\ket{h}}{\bracket{h|h}} =
\frac{\sum_{v,v'}W_{v,v'}\frac{\bra{v}\hat{O}\ket{v'}}{\bracket{v|v'}}}{\sum_{v,v'}W_{v,v'}},
\end{equation}
can be interpreted as an ensemble average of the estimator
$\bra{v}\hat{O}\ket{v'}/\bracket{v|v'}$
in a fluctuating gas of valence bond loops.~\cite{Sandvik05,Beach06}
Over the range $g < g_M$, the bond amplitudes are all
strictly positive and thus the sampling weight
\begin{equation}
W_{v,v'} = \bracket{v|v'} \biggl[ \prod h(\vec{r}) \biggr]\biggl[ \prod h(\vec{r}') \biggr]
\end{equation}
has no sign problem associated with it.
Numerical evaluation of the RVB wavefunction is 
carried out using a worm algorithm~\cite{Prokofev01} 
adapted to the valence bond loop gas.~\cite{Beach07c}

\begin{figure}
\includegraphics{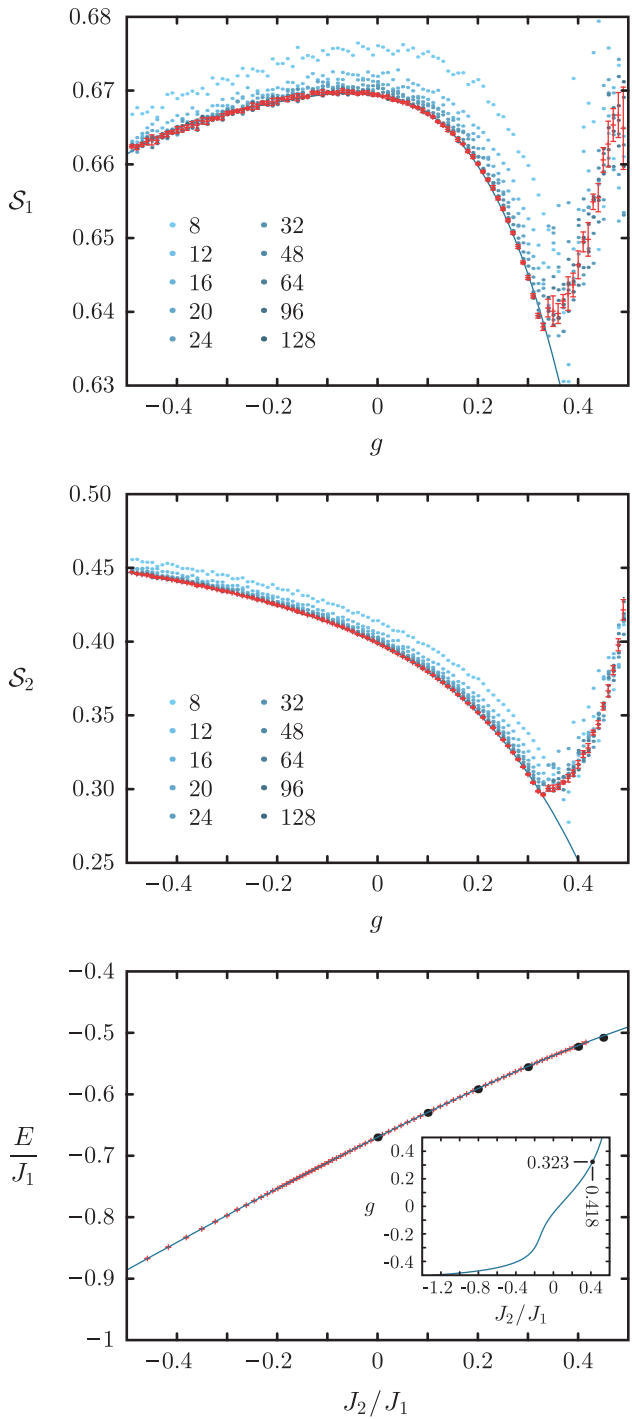}
\caption{\label{FIG:en}
(Top) The expectation value of the NN
spin correlations
$\mathcal{S}_1 = -(1/2N)\sum_{\vec{r}} \bracket{\vec{S}_{\vec{r}} \cdot \vec{S}_{\vec{r} + (1,0)}}$
is plotted as a function of the variational parameter $g$. 
Solid circles (blue) represent data computed
for a particular $L \times L$ system; see the legend. 
Error bars (red) denote the $L \to \infty$ extrapolation.
(Middle) The NNN spin correlations
$\mathcal{S}_2 = (1/2N)\sum_{\vec{r}} \bracket{\vec{S}_{\vec{r}} \cdot \vec{S}_{\vec{r} + (1,1)}}$
are plotted in the same way.
(Bottom)
The energy density
$E/J_1 = -\mathcal{S}_1 + (J_2/J_1)\mathcal{S}_2$
in the thermodynamic limit is compared to
estimates (black dots)
due to Gochev.~\cite{Gochev94}
The inset shows the $g$ that minimizes the variational energy.
}
\end{figure}

Figure~\ref{FIG:en} shows the NN and NNN spin correlations
computed for finite lattices as a function of $g$. These data are extrapolated 
to the thermodynamic limit by assuming $O(L^{-3})$ leading corrections.
A weighted sum of the correlations gives the variational energy:
\begin{equation} \label{EQ:varE}
E(g) = \frac{J_1}{2N}\sum_{\vec{r}}\biggl[ \bracket{\vec{S}_{\vec{r}}\cdot\vec{S}_{\vec{r}+(1,0)}} + 
\frac{J_2}{J_1}\bracket{\vec{S}_{\vec{r}}\cdot\vec{S}_{\vec{r}+(1,1)}} \biggr].
\end{equation}
The optimal value of $g$ is found by solving $E'(g_{\text{min}})=0$.
The dependence of $g_{\text{min}}$ on $J_2/J_1$ is shown as an inset 
in the bottom panel of Fig.~\ref{FIG:en}.
Back substitution of $g_{\text{min}}$ into Eq.~\eqref{EQ:varE} gives
$E(J_2/J_1)$.

As is clear from the top and middle panels of Fig.~\ref{FIG:en}, 
a point of nonanalyticity at $g=g_M$ (or $J_2/J_1 = 0.418$)
separates regions with markedly different behaviour.
In the $L\to\infty$ limit, both the NN and NNN spin correlations 
exhibit a cusp. $E'(g)$ has no roots for $g\ge g_M$.

\begin{figure}
\includegraphics{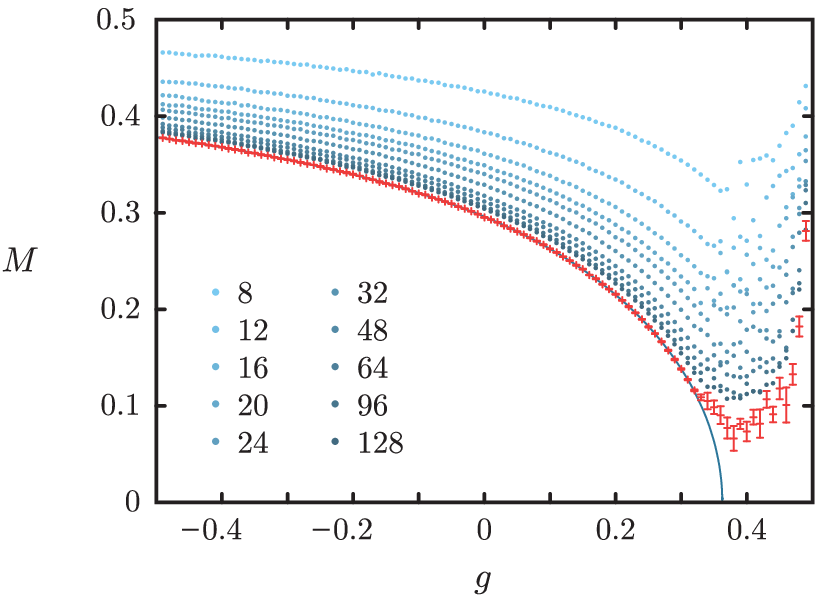}
\includegraphics{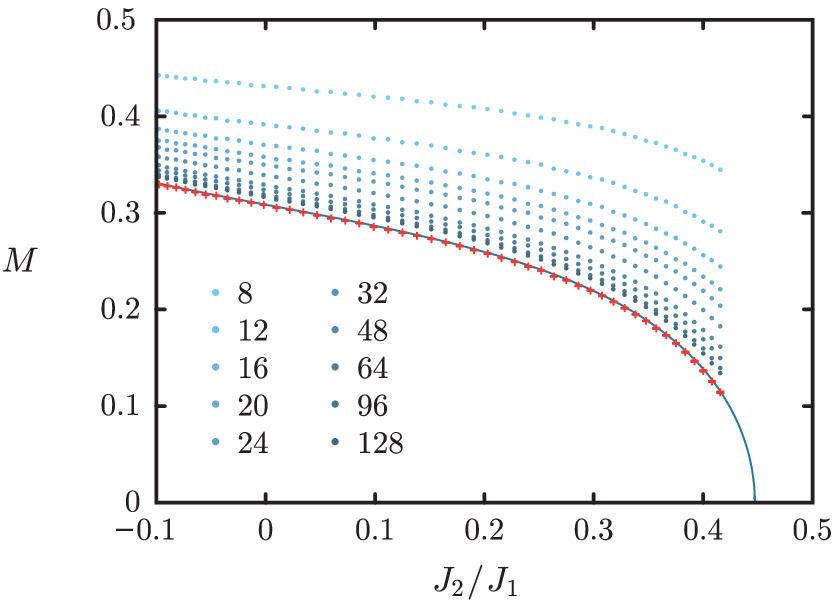}
\caption{\label{FIG:mag}
(Top) The staggered magnetic moment $M$ is plotted as a function of
the variational parameter $g$. The dots (in shades of blue)
represent data computed for a particular $L \times L$ system.
Errorbars (red) indicate the extrapolated $L\to\infty$ value.
The data for $g > g_M$ is unreliable for reasons elucidated in
the main text.
The solid line is a polynomial fit of the $L=\infty$, $g < g_M$ points.
(Bottom) The $g < g_M$ data from the top panel is replotted with
a new horizontal scale. The Marshall sign rule breaks down at 0.418,
and the magnetic order dies at 0.447.
}
\end{figure}

Figure~\ref{FIG:mag} shows the staggered magnetic moment
\begin{equation}
M = \frac{1}{2N}\biggl[\,\sum_{\vec{r},\vec{r}'}(-1)^{\lVert \vec{r}-\vec{r}'\rVert}\bracket{\vec{S}_{\vec{r}}\cdot\vec{S}_{\vec{r}'}}\biggr]^{1/2}
\end{equation}
plotted as both $M(g)$ and $M(J_2/J_1)$. Here, the thermodynamic
limit is acheived with $O(L^{-2})$ scaling. By continuing the trend established in the
region where the Marshall sign rule is obeyed, we estimate that the staggered moment
vanishes continuously at a critical coupling $J_2/J_1 = 0.447$.

\section{Discussion}

The master equation approach applied to the $J_1$--$J_2$ model
leads to an RVB trial wavefunction whose bond amplitudes
depend on a single variational parameter, $g$.
Wherever the Marshall sign rule holds, we are able to compute the properties of 
the RVB state to very high accuracy for large lattices (up to $L=128$ easily on a laptop)
and thus to extrapolate measured values to the thermodynamic limit.
For the NN Heisenberg model ($J_2 = 0$),
the energy and staggered magnetization of
the best variational state (at $g=-0.0484$)
extrapolate to $E=-0.669748$ and $M=0.3086$.
These differ by 0.047\% and 0.52\% from the exact results
$E =  -0.669437(5)$ and $M = 0.3070(3)$ obtained from
quantum Monte Carlo.~\cite{Sandvik97}

As a function of frustration, the antiferromagnetic order dies out quite slowly.
When the Marshall sign rule finally breaks down (at $J_2/J_1 = 0.418$),
the staggered moment has decreased only to $M=0.1114$, about
36\% of its unfrustrated value. This does not appear to be consistent
with a continuous transition.
One point of concensus for this model is that a gapped
state appears around $J_2/J_1 \approx 0.4$. Thus, the fact that
the magnetization is still large near this value suggests 
that the transition is first order. In comparison to other
situations where a N\'{e}el--VBS transition is known
to occur,\cite{Sandvik07a,Beach07a} what we observe here 
is much more like the situation in Ref.~\onlinecite{Beach07a}, 
where the staggered magnetization decreases only modestly 
and then collapses abruptly at a first-order critical point.

Of course, this line of reasoning is not sufficiently rigourous to establish
the order of the transition, and we cannot rule out a deconfined 
quantum critical point. Since there are no exact results on large
lattices, we do not know how well the optimized RVB wavefunction
approximates the true ground state. (For $4\times 4$ the agreement 
is quite good\cite{Lou06}: the overlap is 0.9998 for the unfrustrated
case and 0.996 for $J_2/J_1=0.4$). We have estimated that the magnetic 
order vanishes at $J_2/J_1 = 0.447$ in a continuous scenario, which 
apparently ``overshoots'' the openning of the spin gap at $J_2/J_1\approx 0.4$.
It is hard to say whether these values are truly noncoincident, since their
uncertainties are difficult to quantify. Moreover, the decay of the
staggered magnetization may be artificially slow because of the failure of the RVB
state (which is translationally invariant) to capture
the incipient dimer correlations near the transition.

The sign problem in this model (for $J_2/J_1>0.418$) turns out not 
to be terribly severe. Much more catastrophic is that the master equation 
itself breaks down along with the Marshall sign rule, because of 
the assumption that $h(\vec{r}) \ge 0$ represents the probability of 
finding a bond of type $\vec{r}$. Once any of the amplitudes becomes
negative, the reasoning that lead to Eq.~\eqref{EQ:J1J2mastereq} is no longer
correct. The breakdown could perhaps be avoided if we were to use an exact numerical 
implementation of Eq.~\eqref{EQ:htau} to find the $\tau \to \infty$ limit,
rather than an analytical ansatz. More likely, though, the failure of the master equation
is related to the inability of the RVB state to accommodate bond-bond correlations---except indirectly 
by strengthening the C${}_4$ symmetry of $h(\vec{r})$, as seen in Fig.~\ref{FIG:hr}. 

The master equation approach works remarkably well in guiding our choice of
the RVB bond amplitudes. Where it can be checked ($J_2=0$), the accuracy of the
wavefunction rivals that of unbiased optimizations, but with an enormous
computational saving associated with reducing the number of variational parameters from $N$ to 1.
Including variational parameters for a few additional modes (as described at the end of Sect.~\ref{SECT:frustint})
would improve the accuracy further. In order to handle the most disruptive
effects of frustrating interactions, however, it will be necessary to move
to the next level of approximation and to consider RVB states whose weights factorize
into amplitudes for \emph{pairs} of bonds. Obtaining an analytical master
equation for the two-bond amplitude $h_{ij;kl}$, as we did in this paper for
the single-bond amplitude $h_{ij} = h(\vec{r}_{ij})$, is probably 
not feasible. Nonetheless, for variational calculations, it may be enough to put in by hand some bond-bond 
contribution, \emph{e.g.},
\begin{multline}
h_{ij;kl} = h(\vec{r}_{ij})h(\vec{r}_{kl})\\
\times\bigl[
1 + U\delta(\lVert \vec{r}_{ij}+\vec{r}_{kl}\rVert)\delta(\lVert \vec{r}_{il}+\vec{r}_{kj}\rVert)\bigr],
\end{multline}
that is compatible with the expected VBS pattern (here, the columnar state at large $U$).

\begin{acknowledgments}
The author gives warm thanks to Anders Sandvik and Valeri Kotov for
many stimulating discussions. Financial support was 
provided by the Alexander von Humboldt foundation.
\end{acknowledgments}


\begin{thebibliography}{99}
\bibitem{NoteB} The factorizable form can be justified by a mean field decomposition of the
Heisenberg hamiltonian in terms of bond operators. The amplitudes $h(\vec{r})$ are predicted
to decay as $1/r^{d+1}$; see Ref.~\onlinecite{Beach07b}.
\end{thebibliography}
\end{document}